# Architectural Support for Global Smart Spaces

Alan Dearle[1], Graham Kirby[1], Ron Morrison[1], Andrew McCarthy[1],
Kevin Mullen[1], Yanyan Yang[1], Richard Connor[2], Paula Welen[2], and Andy Wilson[2]

[1] School of Computer Science, University of St Andrews,
North Haugh, St Andrews, Fife KY16 9SS, Scotland
{al, graham, ron, ajm, kevin, yyyang}@dcs.st-and.ac.uk
[2] Department of Computer Science, University of Strathclyde,
Livingstone Tower, 26 Richmond Street, Glasgow G1 1XH, Scotland
{firstname.lastname}@cis.strath.ac.uk

**Abstract.** A GLObal Smart Space (GLOSS) provides support for interaction amongst people, artefacts and places while taking account of both context and movement on a global scale. Crucial to the definition of a GLOSS is the provision of a set of location-aware services that detect, convey, store and exploit location information. We use one of these services, hearsay, to illustrate the implementation dimensions of a GLOSS. The focus of the paper is on both local and global software architecture to support the implementation of such services. The local architecture is based on XML pipelines and is used to construct locationñaware components. The global architecture is based on a hybrid peerñtoñpeer routing scheme and provides the local architectures with the means to communicate in the global context.

## 1 Introduction

The ubiquitous computing paradigm has the goal of providing information and services that are accessible anywhere, at any time and via any device [1]. Within this paradigm, a GLObal Smart Space (GLOSS) provides support for interaction amongst people, artefacts and places while taking account of both context and movement on a global scale.

We are engaged in the construction of a GLOSS as part of the Gloss project [2] within the EC Disappearing Computer Initiative [3]. Our GLOSS framework describes a universe of discourse for understanding global smart spaces. The key concepts are **people**, **artefacts** and **places**. People have **profiles** that define their context. A **conduit** is a particular type of artefact that is associated with a person and is part of, or in communication with, the general (connected) GLOSS fabric. The fabric comprises a globally distributed overlay network of static and mobile **nodes** communicating over various transport mechanisms as appropriate.

Of particular importance to Gloss, given its emphasis on *global* and *spaces* and our interest in location-aware services, is the treatment of **places** and other geo-spatial concepts, unified by the general term **where** (selected to avoid the inevitable connota-tions associated with more precise terms). A number of auxiliary concepts are significant in the GLOSS framework but are not essential for this discussion.

Central to our GLOSS is the provision of *location-aware services* that detect, convey, store and exploit location information. In this paper, in order to illustrate our architecture, we concentrate on one such service: *hearsay*[1]. It is described as a composition of the GLOSS concepts and forms a basis for the design of practical tools to support mobile users in a global context.

The contribution of this paper is a software architecture that facilitates the implementation of location-aware services both locally and globally.

## 2 Region Transition Hypothesis

The tractability of implementing a GLOSS (as opposed to a small-scale, restricted, smart space) hinges on our proposed *region transition hypothesis* which states that:

> As the granularity of the geo-spatial region increases, the total frequency of transitions between regions falls.

Globally, there are millions of transitions occurring every second at a fine granularity: a user in France enters a cafÈ at the same instant that a user in the UK leaves a shop and a user in Ireland goes into a bar. There may be 50 shops, cafÈs and other GLOSS regions in a street, with frequent transitions amongst them, but the street itself has only a few access points (its junctions with other streets) and users move between streets less frequently. Similarly, the overall frequencies of transitions between cities and countries are still lower.

The *region transition hypothesis* has implications for the GLOSS architecture and, in particular, for caching. If servers are organised into a tree, with each server being associated with a smaller geo-spatial region than its parent, data may be effectively cached at the nodes of the tree obviating the need to ship high volumes of data, for example profile data. However, tree architectures are not generally scalableó as the root node is approached, the volumes of traffic increase and some mechanism is required to prevent network saturation close to the root. In Section 5.2 we propose a hybrid overlay network based on peerñtoñpeer (P2P) to address this problem.

## 3 Implementation Dimensions

When engineering a GLOSS, many implementation dimensions must be considered, including: location detection; storage; computation; communication; identity of users and devices; user interface; and format and content of information. Since we concentrate here on location-aware services, we discuss only the first four dimensions.

---

[1] We have also identified Radar and three kinds of Trails as locationñaware services [4]

**Location detection**: Two broad classes of location detection devices are considered: those where the artefact being detected has knowledge of its position; and those where the external fabric has knowledge of the artefactís position.

An example of the former is a GPS device. It is aware of the position of its user but that knowledge is not available in the environment without it being actively propagated. An example of the second is an RF detector that can detect a person or artefact carrying a tag, where the person or artefact has no knowledge of their own position. Hybrids of these location detection mechanisms exist. For example, wireless Ethernet cards in mobile devices may determine their position by triangulation with fixed based stations and conversely may themselves be detected by other base stations.

A GLOSS should be able to exploit whatever location detection technology is available. This means using different technologies in different contexts.

**Storage**: The cost, bandwidth and capacity of storage, and the durability of data stored vary considerably from device to device. Mobile phones are at one end of the spectrum having relatively low volumes of storage with data typically being stored in one placeó the device itself. Server nodes are at the other extreme with high volumes of low cost storage featuring high bandwidth access, which is easily and cheaply replicated. Data resident on servers tends to have better availability characteristics than data resident on hand-held devices.

A GLOSS may store information in the locations most appropriate for the scenario supported. This may necessitate maintaining cached copies of client data on servers in order to: obviate high communication costs to devices; have data available close to where computation will occur; and make data available when devices are disconnected.

**Computation**: There are three possibilities as to where computation may occur in a GLOSS, namely at the client, at the server, and in the network. Computation at the client is usually limited in terms of CPU power, storage capacity, I/O bandwidth and battery capacity. We therefore normally discount the client as a site for heavy computation such as performing complex matching or database queries. However computation at the client is necessary for a number of tasks including: user event notification, the integration of data from devices (such as GPS receivers) and the processing of information sent to the client. None of the limitations of the client exist at the server, making it more suitable for heavy computation. Processing may also take place in the network should computational resources be available there.

The GLOSS infrastructure should make the most appropriate use of whatever resources are available. This often means performing matching and searching tasks on servers and conserving battery life on handheld clients by using them only for the presentation of data. However, the placement of computation is always balanced by the cost of moving data to an appropriate place for computation to occur.

**Communication**: In a GLOSS, a variety of communication mechanisms is required for inter-server, inter-client, and client-server communication. Inter-server communication is dominated by Internet protocols and this seems likely to continue. However, a variety of technologies is available for (mobile) inter-client, and client-server communication including GPRS, SMS, TCP/IP via a modem connection, TCP/IP via wireless ethernet, and Bluetooth. These technologies have quite different characteristics in terms of end-point connections, cost and bandwidth.

## 4 Hearsay

To motivate our design decisions on the software architecture for implementing location-aware services we introduce an example service based on hearsay. Intuitively hearsay provides a means for the user to leave and obtain information specific to particular geoñspatial regions, circumstances and personal preferences. It is the electronic equivalent of being able to leave Post-It notes for users with a particular profile in a particular location. Hearsay can be described in terms of the key concepts as:

```
Where X Profile -> Information
```

That is, hearsay provides a means for the user to send and receive messages that are delivered only when the receiver enters a specific **where** and has a matching profile for the hearsay. Two services are required for hearsay: *insertion* and *delivery*. Hearsay insertion allows a user to insert a message, parameterised by a **where** and a **profile**, into the GLOSS fabric at any point. The **where** describes the geoñspatial region in which the message should be delivered, and the **profile** restricts delivery to individual users with particular interests.

Hearsay delivery is triggered when a user, with a matching profile, is detected entering a **where** that has associated, relevant hearsay attached to it. There are two separate implementation aspects: how the event describing that user movement reaches the hearsay delivery service, and how the hearsay is then delivered to the user.

### 4.1   Hearsay Scenario

To illustrate the hearsay service, consider the scenario where user Anna inserts some hearsay relevant to Bob (about a cafÈ) at a **where** (the street containing the cafÈ), due for delivery when Bob enters the street. The insertion can originate from any GLOSS node. Hearsay is placed (stored) on the GLOSS node that corresponds to the **where** in which the hearsay should be deliveredó if there is such a nodeó or, otherwise, on the GLOSS node whose **where** most closely contains the delivery **where**.

Bob has a PDA that is GPS-enabled, and communicates with the GLOSS fabric via SMS. **Fig 1** shows Bobís PDA sending SMS messages, containing his coordinates an SMS gateway in Brussels. The SMS gateway processes the coordinates data and informs the streetís GLOSS node, identified from the coordinates, that Bob has entered the street. Once Bobís profile is matched, by the streetís GLOSS node, with that of the hearsay, the hearsay is delivered to him.

### 4.2   Implementation of the Hearsay Metaphor

Hearsay insertion is a storage activity. However this begs the question as to where is the best place to store the association between a region and the hearsay information associated with it.

> **Key decision**: Hearsay is stored on servers associated with particular geo-spatial regions. Each server is located within the geo-spatial region with which it is associated.

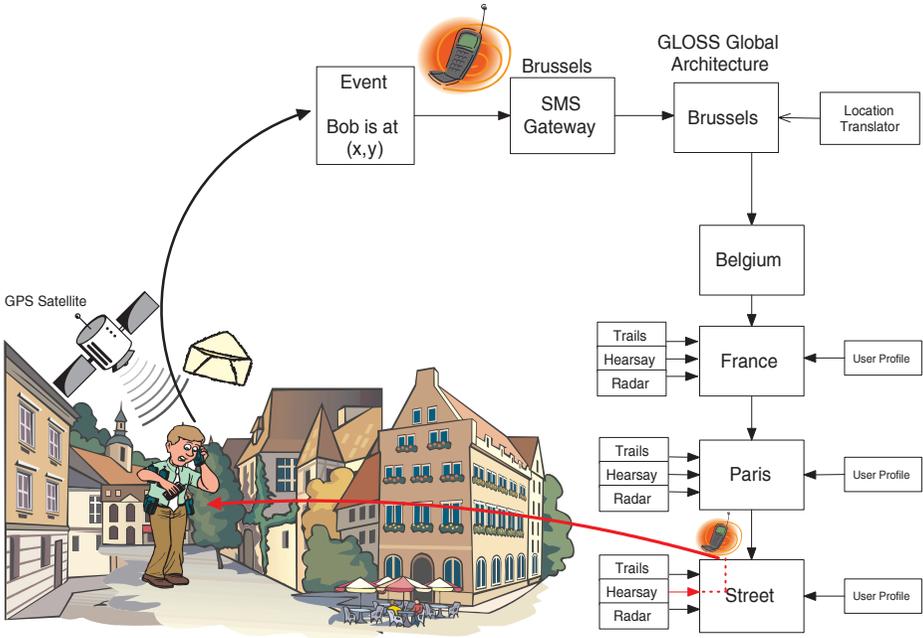

**Fig 1**: Hearsay delivery anywhere in a GLOSS-aware **where**; location detection by user

We believe that this is a good compromise since the alternatives all have serious drawbacks. Storing hearsay on the depositorís node makes it hard to deliver when a user enters a **where**. Using central servers to store hearsay is easy to manage but has limited potential for scalability. If hearsay is stored on a server associated with a particular geo-spatial region, it simplifies adding additional servers thereby making it scalable. Broadcasting or publishing the association on the network is an attractive option to facilitate the discovery of hearsay. Later in this paper we outline how it is possible to build a P2P network that routes messages to an appropriate server.

Hearsay delivery is triggered when a user enters a **where** that has associated, relevant hearsay attached to it. The implementation of hearsay delivery has four technological dimensions:

- where the detection of the userís location takes place
- the communication between where the detection has occurred and where the profile matching takes place
- where the computation matching the userís location to the hearsay takes place
- how the user is notified of the hearsay

The detection of the userís location may occur in a number of places depending on the technological mix. However, the nodes receiving the detection event must do something with itó again there are a number of choices. Each node could, for example, send the event to the userís home node or broadcast the event.

The first choice assumes that each user has a designated *home* node. This is an attractive optionó it scales welló servers can be added on demand as users are added. However, it does not solve the problem of hearsay matchingó it only delays it. When the message is sent to the userís *home* node, another message must be sent elsewhere for the matching to occur. In the second choice, the event may be broadcast and used by whichever services are interested in it.

However some mechanism must to be provided to route messages to those servers that are interested in the events. Communication between where the detection takes place and where the matching occurs can be performed in a number of ways. At one extreme standard Web infrastructure (HTTP) could be used for all inter-node communications; this approach is followed in cooltown [5]. At the other extreme is the use of overlay networks, utilising non-standard protocols. In order to exploit the *region transition hypothesis*, some customised routing mechanism needs to be provided. This could be provided above standard protocols such as TCP or HTTP.

There are three main contenders for where the computation matching of the userís location to the hearsay takes place: on the userís conduit, on the userís home node, or on the node storing the hearsay. Performing matching on the userís conduit is probably a bad choice: it is wasteful of two precious resources, namely network bandwidth and battery life. Furthermore, memory and CPU resources on the conduit are likely to be relatively limited.

Performing matching on the userís home node (if there is one) is also a poor choice since it necessitates all potentially applicable hearsay to be sent across the network. The alternative is to perform matching where hearsay is stored. If the user profile is large (we have empirical evidence to suggest it would not be) the option we have chosen appears to be a poor one. However, we believe that it is not, due to the *region transition hypothesis* described earlier.

> **Key decision**: The computation matching hearsay to the userís location is performed on the nodes storing the hearsay.

The last issue is that once the hearsay service has matched the user and the hearsay, it must send the hearsay to the user. There are many choices for this delivery and therefore some abstraction over the technologies is required to insulate the hearsay services (and other services) from the network.

> **Key decision**: Each userís profile contains information about how to contact that user. This may include alternatives to deal with failure.

## 5 GLOSS Architecture

The scenario highlights two points:
- A local architecture is required to permit GLOSS clients and servers to be constructed. Both clients and servers are essentially data-flow architectures. Data arrives at a userís device from several sourcesó communications devices, position servers etc.ó and is passed to elements for processing. These might include displaying information to a user, sending it to another device, or storing it for future presentation or processing. Similarly on a server, data arriving from several de-vices is processed and delivered to other devices for further processing or presentation to the user.
- A global architecture is required to mediate the global flow of information between clients and servers.

To permit GLOSS services to be constructed, we have designed an architecture based on XML pipelines, with XML events flowing between pipeline components, both locally within an individual client or server node, and globally between nodes.

### 5.1    Local Architecture

Our GLOSS architecture is designed with the following motivations:
- to abstract over any particular technology
- to make GLOSS components independent of each other
- to allow components to be assembled into GLOSS applications

The local architecture is based on XML pipelines, and XML event buses that allow incoming events to be delivered to multiple downstream components. To abstract over location, GLOSS clients transparently pass messages to servers and vice-versa by plugging appropriate components into the pipelines and event buses of the client and server.

Data enters an individual pipeline from various devices such as GPS and GSM devices, and from pipelines running on other nodes. The data takes various formats, including device-specific strings (e.g. NMEA strings emitted by a GPS device), XML fragments, and typed objects. Adapter components convert between these formats as necessary. Each hardware device has a wrapper component that makes it usable as a pipeline component. Other components perform filtering (e.g. transmitting user-location events only when the distance moved exceeds a certain threshold), buffering, communication with other pipelines, and so on. Abstract interfaces define the minimum required functionality of a pipeline component.

The components and interfaces described above achieve the first two objectives. In order to manage collections of component instances we introduce another abstraction– the *assembly*. This is a collection of components that are linked together via pipes and buses.

To accommodate the scenario, a pipeline is assembled on Bob's conduit containing:
- a GPS device
- an adapter to convert GPS strings into objects
- a position threshold filter to reduce user-location events to the desired granularity
- an event bus
- a hearsay user interface tool and an SMS device, both attached to the event bus

This assembly is shown on the left in **Fig 2**:

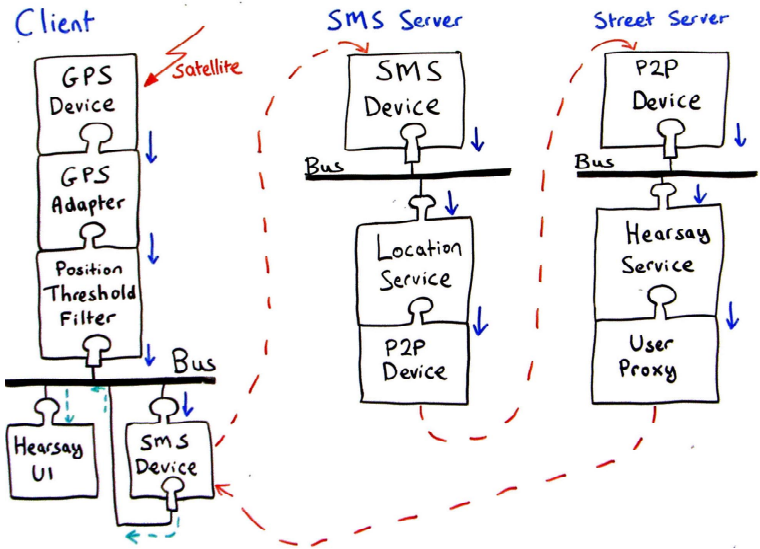

**Fig 2**: Pipeline assemblies to support hearsay delivery

In the scenario, the GPS device passes events via the adapter to the event bus, which passes them to the SMS device. This sends the events to Bobís SMS server located in Brussels.

On the SMS server, another pipeline is assembled, capable of processing incoming SMS events. In this instance of the architecture, a location service component is attached to the event bus. It passes location information to a P2P component which results in a message being passed to a street server located in the same street as Bob.

A third pipeline is assembled on the street server. It contains a P2P component to receive incoming object events, which flow into an event bus, which in turn has a hearsay service component plugged into it. The hearsay service determines that Bob is in the street (determined by the event sent to it), and that a cafÈ recommended by Anna is close by (the hearsay). A message needs to be sent to Bob to inform him of this fact. The message could be sent to Bob using a variety of mechanisms. To abstract over these mechanisms, a proxy component for Bob is attached to the pipeline. This component sends a message to Bob using the most appropriate technologyó in this case, an SMS message.

The SMS device on Bobís conduit receives the SMS message, where it is passed to the event bus and from there to the hearsay user interface tool, which presents the hearsay to Bob.

## 5.2   Global Architecture

We have described a local architecture for GLOSS conduits and servers. Next we briefly outline how these components may be organized into a global architecture for smart spaces. Tree architectures are not scalableó as the root node is approached, the volumes of traffic increase [6]. This problem may be addressed using peerñtoñpeer architectures. However, most P2P architectures only search a limited region and are therefore not immediately applicable in this context where messages require routing to a specific destination. For example, in the hearsay example, a SMS location message must be routed to an appropriate hearsay server to enable matching to occur. However, it may be injected into a GLOSS network a considerable distance (in terms of both hops and geo-spatial proximity) from a server storing hearsay.

The network supporting the GLOSS infrastructure must therefore provide some mechanism to route messages to an appropriate location. Hybrid architectures offer the opportunity to tailor topologies and protocols in such a way that locality knowledge may be exploited. In [6], Carzaniga et al state that an acyclic peer-to-peer architecture with subscription forwarding appears to scale well and predictably under all circumstances and is likely to represent a good choice to cover a wide variety of scenarios. Consequently, we are currently exploring a global peer-to-peer architecture consisting of a hierarchy of peers such as that shown in **Fig 3**.

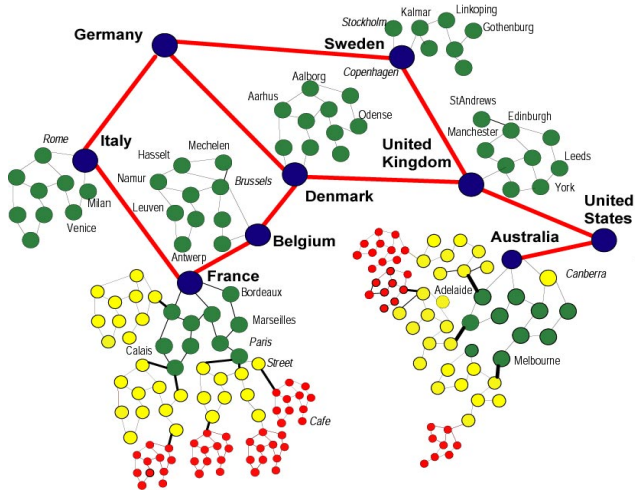

**Fig 3**: Hybrid global architecture

In addition to avoiding network bottlenecks, a hierarchy of peers is well suited to the geo-spatial nature of GLOSS queries. As shown in **Fig 3**, it is relatively easy to partition the world recursively into non-overlapping regions. These may be mapped on social, economic or organisational boundaries. An advantage of the hybrid routing scheme is flexibility in the Key Decision given in Section 4.2. Now servers need not be co-located with the geo-spatial region that they representó the only necessity is that peers understand the peering relationships. A final benefit of this architecture is that it is capable of evolving to cope with stress placed upon it.

Partitioning the network into a hierarchy of peers makes the routing of messages to appropriate locations with the GLOSS infrastructure relatively straightforward.

In the scenario, Bobís conduit relays his position to an SMS server located in Brussels. As described above, the GLOSS object pipeline on that server examines location and determines that it must be sent elsewhere. Since the SMS server is part of a P2P network, it does not require a final destination address for the message. However, it must determine if the message should be sent to its children, its parent, broadcast to its peers or dealt with locally. This calculation may be performed by comparing the geo-spatial location of the user with that being managed by the server. In the case of the message about Bobís location, the location in the message is about a location in Paris so it is passed to the Brussels server which either sends it directly to the Paris server (if it has knowledge of this server and the geo-spatial region it manages) or broadcasts it to its peers. When the Paris server receives this message, it is passed down the hierarchy until it reaches the street server where the computation matching the hearsay to Bob can occur.

The hierarchy of peers dovetails well with the region-transition hypothesis. The hierarchy gives a nested set of locations in which Bobís profile may be cached (stored). Assuming Bobís profile is stored at his *home* server, when Bob is first detected in France, a request for Bobís profile will be propagated up the network and down to Bobís home in a manner similar to that described above for hearsay messages. The reply may be cached in servers at multiple levels in the network to obviate the need to repeatedly fetch this information. A similar technique under-pins the replicated document cache used in Freenet [7]. The use of different cache policies at different levels in the hierarchy is likely to be beneficial. For example, leaf servers (where change is fast) might only cache data for a short period but servers close to the root (where change is slow) might cache data for longer periods.

## 6 Related Work

P2P architectures, for example Gnutella [8] and Kazaa [9], have recently become popular for file sharing applications. Many of these architectures depend on searching a limited network centric horizon of machines governed by a number of network hops from the point of query. Such architectures are not perfectly suited to an environment in which messages require routing to a specific destination. A number of P2P systems that perform routing to some destination, such as Pastry [10] and OceanStore [11], have recently been described in the literature. These systems are loosely based on a routing algorithm by Plaxton [12] and perform application level routing on an overlay network to deliver a scalable decentralized architecture for locating objects. In these systems the location of objects (or knowledge of where they are located) is determined by a cryptographic hashing function. After careful examination of this approach, we believe it unsuitable for the geo-spatial location mechanisms described in this paper.

Some local area P2P systems have commonality with and are complementary to the work described in this paper. For example, the STEAM system [13] from our Gloss partners provides a local area ad-hoc network which provides support for typed events and filters over them. We see this work as being an enabling technology to support proximity based group communication, which in turn will feed the global GLOSS architecture described in this paper.

One of the most closely related projects to Gloss is the HP cooltown project [5]. Unlike Gloss, cooltown is based on Web services and assumes that everything has a Web presence. The cooltown approach, unlike our own, is somewhat anti-middleware ñ instead it builds on pure Web and XML technology. However, it shares many similarities with Gloss in that cooltown aims to ìbridge the physical and online worlds by bringing the benefits of web services to the physical environmentî.

The Gaia project from University of Illinois at Urbana-Champaign aims to ìbring the functionality of an operating system to physical spacesî [14]. This project provides a set of services including an event service and a context service and integrates it with a middleware operating system. Like Gloss this is a middleware approach, however, it focuses on room level rather than global interactions.

## 7 Status of Implementation

At the time of writing, instances of the GLOSS pipeline architecture are running on a mobile device (an iPAQ using the J9 VM from IBM [15]) and a server. A pipeline running on the iPAQ integrates a GPS device, a Nokia GSM phone card and a prototype hearsay user-interface. We have developed Java interfaces to the GPS device which supplies the pipeline with location information, and to the GSM device capable of sending and receiving SMS messages. The server pipeline contains an SMS device which supplies the location information in an XML format to a Unix based Web service supporting a variety of queries over the geo-spatial data.

We are experimenting with a variety of algorithms for global message routing running in a simulator. We plan to integrate these with the local pipeline architectures once we better understand the consequences of the many design decisions that affect the overall system efficiency.

## 8 Conclusions

We have outlined a framework for the description of Global Smart Spaces that supports interaction amongst people, artefacts and places while taking account of both context and movement on a global scale. The essence of the paper is the introduction of a software architecture that facilitates the implementation of locationñaware services such as hearsay within the framework. For scalability we suggest how the software architecture may be applied to provide location-aware services in the global context.

## 9 Acknowledgements

prototype. The work was also supported by EPSRC grant GR/M78403/GR/M76225, ì Supporting Internet Computation in Arbitrary Geographical Locationsî.

## 10 References

1.  Weiser M. The Computer for the 21st Century. Scientific American 1991; September:94-104
2.  Global Smart Spaces. EC 5th Framework Programme IST-2000-26070. 2000. http://www.gloss.cs.strath.ac.uk/
3.  The Disappearing Computer Initiative. Future and Emerging Technologies Activity, EC 5th Framework Programme, 2000. http://www.disappearing-computer.net/
4.  Munro A., Welen P., Wilson A. Interaction Archetypes. GLOSS Consortium Report D4, 2001
5.  Hewlett-Packard. cooltown. 2002. http://www.cooltown.hp.com/cooltownhome/index.asp
6.  Carzaniga A., Rosenblum D.S., Wolf A.L. Design and Evaluation of a Wide Area Notification Service. ACM Transactions on Computer Systems 2001; 19,3:332-383
7.  Clarke I., Miller S.G., Hong T.W., Sandberg O., Wiley B. Protecting Free Expression Online with Freenet. IEEE Internet Computing 2002; 6,1:40-49
8.  Kan G. Chapter 8: Gnutella. In: A. Oram (ed) Peer-to-Peer: Harnessing the Power of Disruptive Technologies. O'Reilly, 2001
9.  Kazaa. 2002. http://www.kazaa.com/
10. Rowstron A.I.T., Druschel P. Pastry: Scalable, Decentralized Object Location, and Routing for Large-Scale Peer-to-Peer Systems. In: Lecture Notes in Computer Science 2218. Springer, 2001, pp 329-350
11. Rhea S., Wells C., Eaton P.R., Geels D., Zhao B.Y., Weatherspoon H., Kubiatowicz J. Maintenance-Free Global Data Storage. IEEE Internet Computing 2001; 5,5:40-49
12. Plaxton C.G., Rajaraman R., Richa A.W. Accessing Nearby Copies of Replicated Objects in a Distributed Environment. In: Proc. 9th Annual ACM Symposium on Parallel Algorithms and Architectures (SPAA '97), Newport, RI, USA, 1997, pp 311-320
13. Meier R. STEAM. 2000. http://www.dsg.cs.tcd.ie/~meierr/pres/pdfs/EventWorkshop_IntroSteam.pdf
14. University of Illinois at Urbana-Champaign. GAIA: Active Spaces for Ubiquitous Computing. 2002. http://devius.cs.uiuc.edu/gaia/
15. IBM. J9 Virtual Machine. 2002. http://www.ibm.com/software/pervasive/products/wsdd/